\documentclass[fleqn,10pt]{wlscirep}

\title{A versatile class of prototype dynamical systems for 
       complex bifurcation cascades of limit cycles}

\author[1]{Bulcs\'u S\'andor}
\author[1,*]{Claudius Gros}
\affil[1]{Institute for Theoretical Physics, Goethe University Frankfurt, 
	Frankfurt am Main, 60438, Germany}

\affil[*]{gros@itp.uni-frankfurt.de}

\graphicspath{{eps/}}
\DeclareMathOperator{\tr}{tr}

\begin{abstract}
We introduce a versatile class of prototype dynamical
systems for the study of complex bifurcation cascades
of limit cycles, including bifurcations breaking 
spontaneously a symmetry of the system, period doubling
bifurcations and transitions to chaos induced by
sequences of limit cycle bifurcations. The prototype 
system consist of a $2d$-dimensional dynamical system with 
friction forces $f(V(\mathbf{x}))$ functionally dependent 
exclusively on the mechanical potential $V(\mathbf{x})$,
which is typically characterized, here, by a finite number 
of local minima.\newline
We present examples for $d=1,2$ and simple polynomial
friction forces $f(V)$, where the zeros of $f(V)$ regulate
the relative importance of energy uptake and dissipation
respectively, serving as bifurcation parameters. Starting
from simple Hopf- and homoclinic bifurcations, complex
sequences of limit cycle bifurcation are observed when
energy uptake gains progressively in importance.
\end{abstract}
\begin{document}
\flushbottom
\maketitle

%%%%%%%%%%%%%%%%%%%%%%%%%%%%%%%%%%%%%%%%%%%%%%%%
\section*{Introduction}\label{sec:introduction}
%%%%%%%%%%%%%%%%%%%%%%%%%%%%%%%%%%%%%%%%%%%%%%%%

The notion of a prototypical dynamical system is 
used whenever a certain system allows to study a
certain relevant phenomenon, being at the same time
simple enough that it is amendable for numerical      
and (at least partial) analytic investigations,
without showing additional properties which could 
obfuscate the study of the prime phenomenon of interest
\cite{cymbalyuk2005neuron,uccar2003chaotic,krupa2008mixed}.
Additionally, their dynamical behavior can often be
understood in terms of general concepts, such as energy 
balance, symmetry breaking, etc.

Examples of prototype systems are the normal forms
of standard bifurcation analysis 
\cite{gros2013complex,chow1994normal}
and classical systems, like the Van der Pol oscillator 
\cite{gros2013complex} or the Lorenz model 
\cite{lorenz1963deterministic}, which have been of
central importance for the development dynamical
system theory. As an example we consider the
Li\'enard equation
\begin{equation}
\ddot{x}+f(x)\dot{x}+g(x)\ =\ 0~,
\label{eq:lienard}
\end{equation}
a generic adaptive mechanical system, which
includes the Van der Pol oscillator and the
Takens-Bogdanov system 
\cite{takens1974singularities,bogdanov1975versal}.
The periodically forced extended Li\'enard systems 
with a double-well potential have also been studied by 
many authors (see e.g. the double-well Duffing oscillator 
\cite{venkatesan1997bifurcation, yamaguchi1989static, li2006chaos}).

In this paper we propose a new class of autonomous  
Li\'enard-type systems, which allows to study cascades 
of limit cycle bifurcations using a bifurcation parameter 
controlling directly the balance between energy dissipation 
and uptake, and hence the underlying physical driving 
mechanism.
Though there are a range of construction methods for dynamical 
systems in the literature (see e.g.\ 
\cite{sandstede1997constructing,deng1994constructing,qi2005four}), 
they tend however to assume a somewhat abstract view, such as 
starting from implicitly defined manifolds, or involve 
mathematical tools accessible only to researchers with
an in-depth math training. In contrast to these methods, we 
provide here a mechanistic design procedure, based on 
the construction of attractors by the use of 
potentials, a concept often used in many 
interdisciplinary fields (e.g. in modeling cardiovascular 
systems \cite{regan2012dynamical} or for solving  
optimization problems \cite{ercsey-ravasz2011optimization}), 
making it easily accessible and implementable for other 
scientific communities (such as neuroscience, biology etc.) as well.

%%%%%%%%%%%%%%%%%%%%%%%%%%%%%%%%%%%%%%%%%%%%%%%%%%%%%%%
\begin{figure}[t]
\centering
\includegraphics[height=0.60\textwidth]{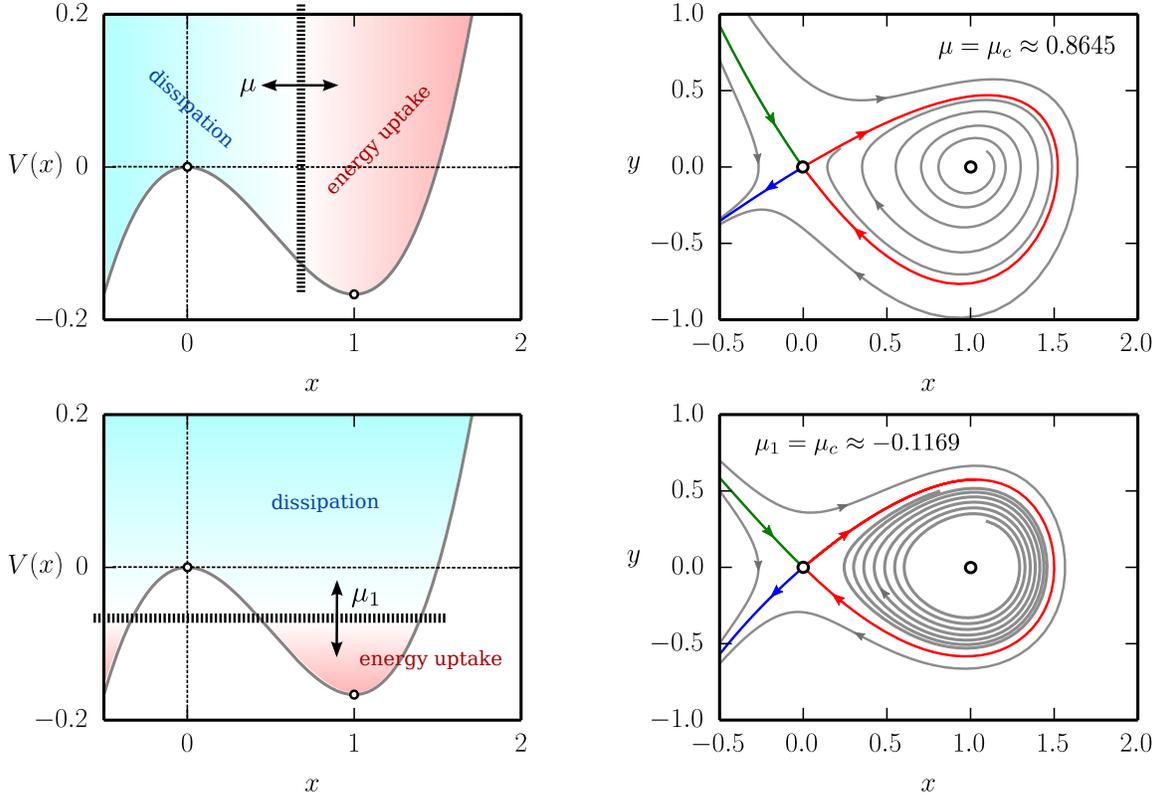}
\caption{The Bogdanov-Takens system (\ref{eq:bt_newton}) 
with the potential function $V(x)=x^3/3-x^2/2$ and a friction
term $x-\mu$ (\textit{top row}), and its generalization 
(\ref{eq:prototype_system}) to a friction term $\mu_1-V(x)$
(\textit{bottom row}), compare 
Eq.~(\ref{eq:friction_functions}).
\textit{Left column}: The potential function together
with the color-coded regions of respectively energy dissipation
and uptake, compare Eq.~(\ref{eq:bt_energy}).
\textit{Right column}: The phase planes at the respective homoclinic 
bifurcation points, with the respective unstable foci and the saddles
by open circles. The green and blue trajectories are the stable and 
unstable manifolds and the red trajectory corresponds to the homoclinic 
loop.}
\label{fig:Bogdanov-Takens}
\end{figure}
%%%%%%%%%%%%%%%%%%%%%%%%%%%%%%%%%%%%%%%%%%%%%%%%%%%%%%%

As an introductory example for the role of balance between
energy uptake and dissipation in both local and global
bifurcations we reconsider the Bogdanov-Takens system,
\begin{equation}
\ddot{x} = (x-\mu)\dot{x}-V'(x), \qquad\quad
\begin{aligned}
\dot{x} &= y\\
\dot{y} &= (x-\mu)y-V'(x)~,
\end{aligned}
\label{eq:bt_newton}
\end{equation}
which is often used as a prototype system for homoclinic 
bifurcations \cite{gros2013complex}. Here, the mechanical
potential is a third order polynomial, as illustrated
in Fig.~\ref{fig:Bogdanov-Takens}. The friction force is
directly proportional to the velocity $y$ and fixpoints
of (\ref{eq:bt_newton}) correspond hence to the
minima and the maxima of the potential $V(x)$. 

The dynamics of the Bogdanov-Takens system is controlled 
by the parameter $\mu$, defining, 
in terms of the mechanical energy $E$, the regions of 
dissipation and energy uptake in the potential valley,
\begin{equation}
\dot E = (x-\mu)y^2, \qquad\quad
E = \frac{y^2}{2} + V(x)~,
\label{eq:bt_energy}
\end{equation}
compare Fig.~\ref{fig:Bogdanov-Takens}. The region $x>\mu$
of energy uptake increases when the bifurcation parameter $\mu$ 
is decreased, leading to two consecutive transitions. Initially
the potential minimum becomes repelling, undergoing a supercritical 
Hopf bifurcation and a stable limit cycle  emerges. Decreasing $\mu$ 
further the extension of the limit cycle increases, merging
at $\mu_c$ with the stable and unstable manifolds of the saddle,
resulting in a homoclinic bifurcation.
%%%%%%%%%%%%%%%%%%%%%%%%%%%%%%%%%%%%%%%%%%%%%%%%%%%%%%%
\begin{figure}[t!]
\centering
\includegraphics[height=0.9\textwidth]{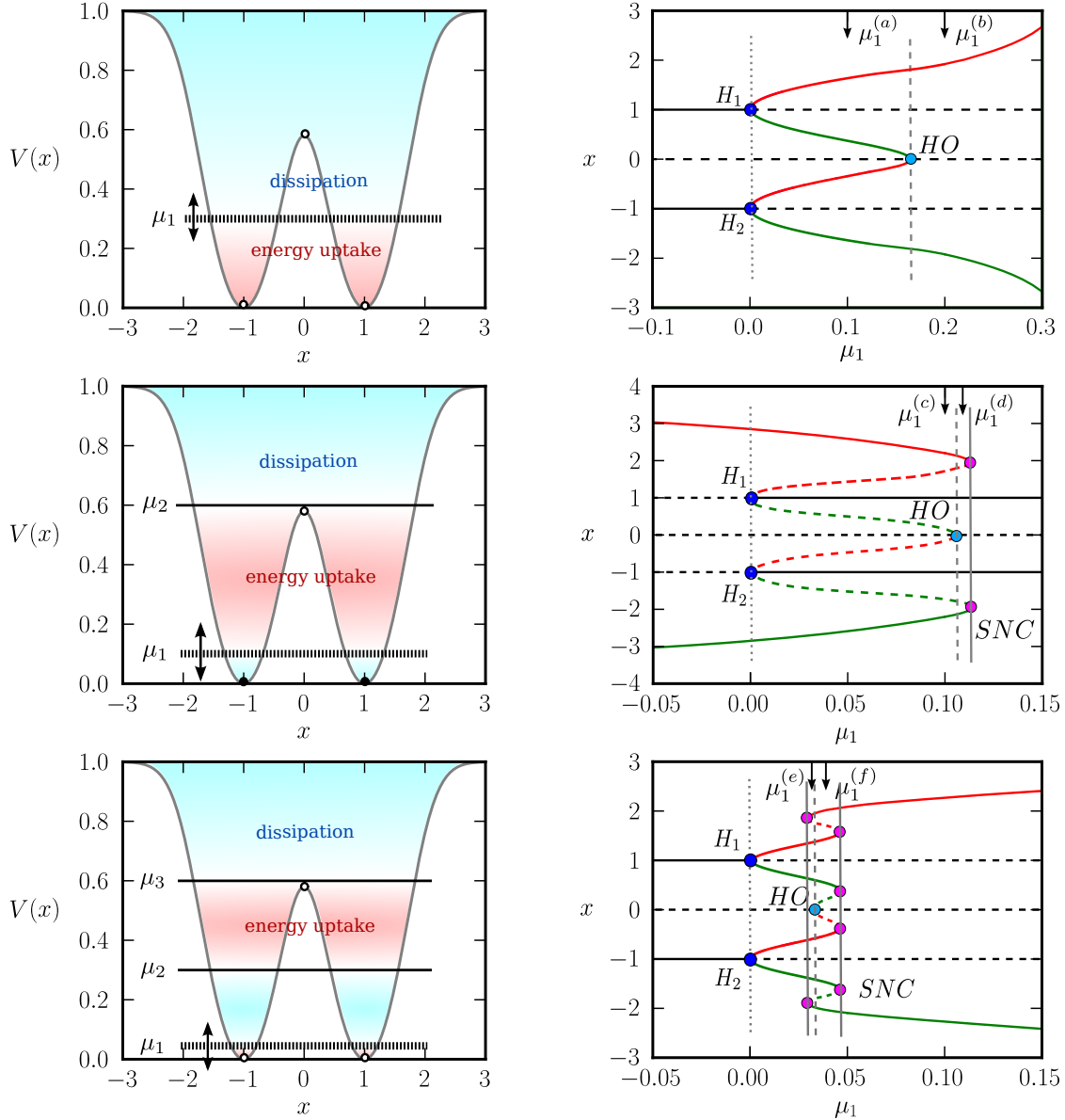}
\caption{\textit{Left column}: Double well 
potential, as defined by Eq.~(\ref{eq:V_predefined}) and
with $x_{1,2}=\pm1$, $V_{1,2}=0$, $z_{1,2}=1$ and 
$p_{1,2}=1$. The regions of energy dissipation $\dot E<0$ 
and uptake $\dot E>0$ are color coded. For the friction 
functions (\ref{eq:friction_functions}) we 
used $f_1$, with $\alpha=1$ (\textit{top row}), 
$f_2$, with $\mu_2=0.6$ and $\alpha=5$ (\textit{middle row}), 
and $f_3$, with $\mu_2=0.3$, $\mu_3=0.6$ and $\alpha=5$ 
(\textit{bottom row}).
\textit{Right column}: Bifurcation diagrams of the respective
generalized Li\'enard systems (\ref{eq:prototype_system}), 
as a function of $\mu_1$. All other $\mu_i$ (when present) are 
kept constant. Stable/unstable fixpoints or limit cycles are 
denoted by continuous/dashed curves respectively. Black lines 
are fixpoint lines, while the maximal/minimal amplitude of $x$ 
in a cycle is denoted with red/green color. 
$H$ points denote Hopf-bifurcations, $HO$ corresponds to 
homoclinic bifurcations of a saddle, $SNC$ points denote 
saddle node bifurcations of limit cycles.
The dotted, dashed and continuous vertical gray lines 
are just guides for the eyes.
}
\label{fig:bifucations_V_diagrams}
\end{figure}
%%%%%%%%%%%%%%%%%%%%%%%%%%%%%%%%%%%%%%%%%%%%%%%%%%%%%%%

%%%%%%%%%%%%%%%%%%%%%%%%%%%%%%%%%%%%%%%%%%%%%%%%
\section*{Results}\label{sec:results}
%%%%%%%%%%%%%%%%%%%%%%%%%%%%%%%%%%%%%%%%%%%%%%%%

The key mechanism leading to a bifurcation in the
Bogdanov-Takens systems is the availability of a parameter
allowing to change the balance between energy uptake
and energy dissipation along limit cycles. Our aim is
to generalize this idea to the case of mechanical systems
characterized by an arbitrary number of potential minima.
For this purpose we consider with
\begin{equation}
\begin{aligned}
\dot{\mathbf{x}} &= \mathbf{y}\\
\dot{\mathbf{y}} &= f(V(\mathbf{x}))\mathbf{y}-\nabla V(\mathbf{x})
\end{aligned}
\qquad\qquad \dot E\,=\,f(V(\mathbf{x}))\,\mathbf{y}^2
\label{eq:prototype_system}
\end{equation}
a $2d-$dimensional system, with $d$ spatial coordinates and with 
friction forces depending via $f(V(\mathbf{x}))$ functionally 
only on the mechanical potential $V(\mathbf{x})$, allowing a 
fine-tuned control of the energy dissipation and uptake around 
the respective potential minima. A well known example of the 
system of type (\ref{eq:prototype_system}) is the 
Van der Pol oscillator
\begin{equation}
\ddot x-\epsilon\left(1-x^2\right)\dot x+x=0,
\qquad\quad f(V) = \epsilon(1-2V),
\qquad\quad V(x) = \frac{x^2}{2}~,
\label{eq:Van_der_Pol}
\end{equation}
for which the regions of energy uptake and dissipation
remain fixed, with $\epsilon$ regulating the
overall influence of the velocity-dependent force.

The simplest generic class of friction functions $f(V)$
entering (\ref{eq:prototype_system})
are polynomial:
\begin{equation}
f_1(V) = -\alpha(V-\mu_1),
\qquad
f_2(V) = -\alpha(V-\mu_1)(V-\mu_2),
\qquad
f_3(V) = -\alpha(V-\mu_1)(V-\mu_2)(V-\mu_3),
\label{eq:friction_functions}
\end{equation}
where $\alpha$ regulates the overall strength of the 
friction and where the individual $\mu_1<\mu_2<\mu_3$ are 
the respective zeros, the points at which dissipation changes 
to anti-dissipation and vice-versa, compare Fig.~\ref{fig:bifucations_V_diagrams}.

When using $f_1(V)$ and the mechanical potential $V(x)=x^3/3-x^2/2$, 
the resulting flow in phase space is equivalent to the one of the 
Bogdanov-Takens system (\ref{eq:bt_newton}), as shown in 
Fig.~\ref{fig:Bogdanov-Takens}. 

%%%%%%%%%%%%%%%%%%%%%%%%%%%%%%%%%%%%%%%%%%%%%%%%%%%%%%%
\begin{figure}[t!]
\centering
\includegraphics[height=0.9\textwidth]{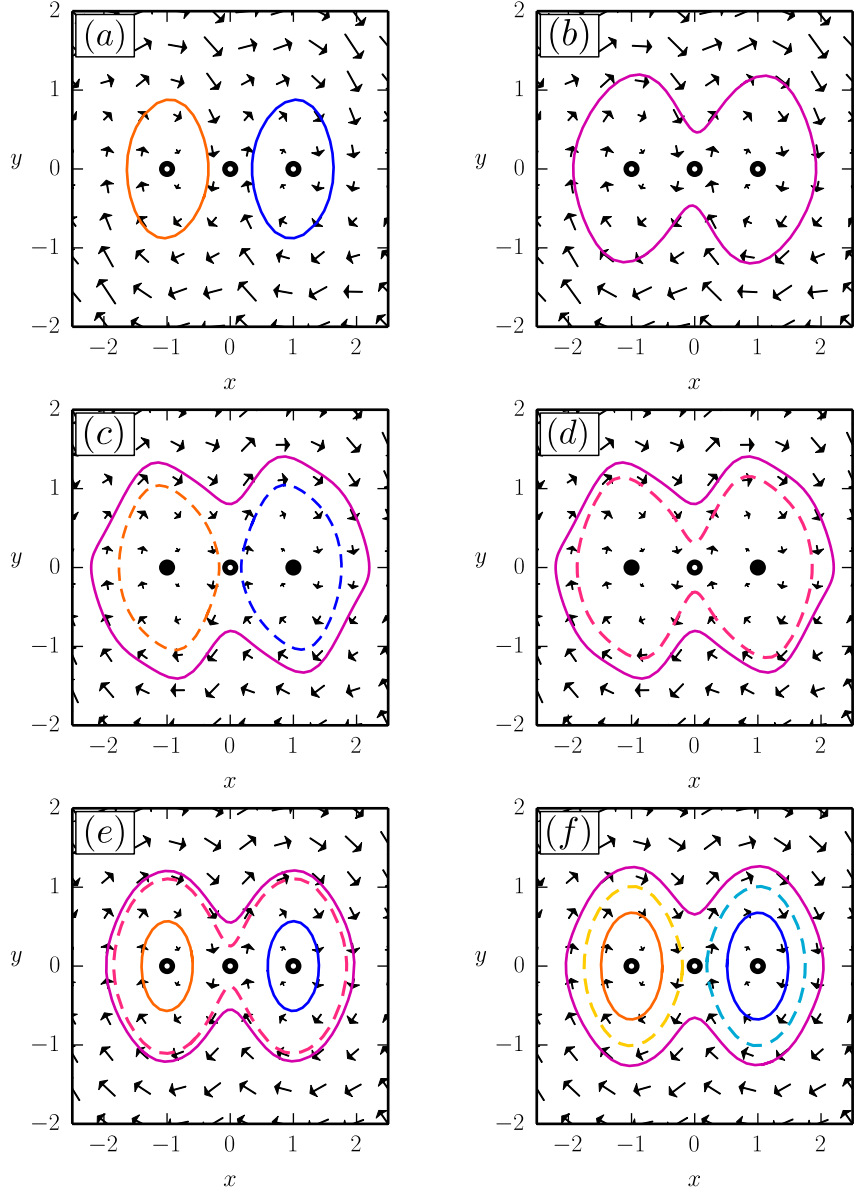}
\caption{Flow diagrams for the systems presented in 
Fig.~\ref{fig:bifucations_V_diagrams}, using respectively linear / quadratic /
cubic friction functions $f_1(V)$ / $f_2(V)$ / $f_3(V)$ 
(\textit{top/middle/bottom row}).
The values $\mu_1^{(a/b)}=0.1/0.2$, 
           $\mu_1^{(c/d)}=0.1/0.11$ and
           $\mu_1^{(e/f)}=0.032/0.04$ 
for the respective $\mu_1$ are indicated by arrows in the 
corresponding bifurcation diagrams in 
Fig.~\ref{fig:bifucations_V_diagrams}.
}
\label{fig:bifucations_flow}
\end{figure}
%%%%%%%%%%%%%%%%%%%%%%%%%%%%%%%%%%%%%%%%%%%%%%%%%%%%%%%
%------------------------------------------------------
\subsection*{Generalized mechanical potentials}
\label{sec:general_potential_well}
%------------------------------------------------------
We are interested in using (\ref{eq:prototype_system}) as 
prototype dynamical systems, especially for the case of 
non-trivial mechanical potentials $V(\mathbf{x})$ having an 
arbitrary number $M$ of local minima. One could in principle
consider higher-order polynomials for this purpose, however 
they do not allow to control the overall height of the 
potential and the relative width of the local minimal in
as simple fashion.

For this purpose we use throughout this study potential 
functions of the kind
\begin{equation}
V(\mathbf{x})=\prod_n \left(g_n(\mathbf{x}-\mathbf{x}_n)+\frac{V_n}{p_n}\right) ,
\qquad\quad
g_n(\mathbf{z})=\tanh(\mathbf{z}^2/{z_n}^2)~,
\label{eq:V_predefined}
\end{equation}
where the $z_n>0$ determine the half-width of the respective
local minima and where the $p_n$ satisfy the self-consistent 
condition:
\begin{equation}
p_{n}=\prod_{m\neq n} \left(g_n(\mathbf{x}_n-\mathbf{x}_m)+\frac{V_m}{p_m}\right),
\qquad\quad
V(\mathbf{x}_n) = \frac{V_n}{p_n}\prod_{m\neq n} 
\left(g_n(\mathbf{x}_n-\mathbf{x}_m)+\frac{V_m}{p_m}\right) = V_n~,
\label{eq:V_predefined_p_n}
\end{equation}
since $g(0)=0$. For deep minima, with 
$(z_i+z_j)\ll\left|\mathbf{x}_i-\mathbf{x}_j\right|$, 
the positions and the heights of the local minima 
are close to $\mathbf{x}_n$ and $V_n$ respectively.
We found, that a relative accuracy of $10^{-2}$ for 
$V_n$ can already be achieved in general after three or
four iterations.

%------------------------------------------------------
\subsection*{Limit cycle bifurcation cascades}
\label{sec:bifurcations_cascades}
%------------------------------------------------------

The system of type (\ref{eq:prototype_system}) allows to 
describe complex cascades of limit cycle bifurcations and
in Fig.~\ref{fig:bifucations_V_diagrams} we show some 
illustrative examples using a symmetric double-well potential 
and linear / quadratic / cubic friction functions 
$f_1(V)$ / $f_2(V)$ / $f_3(V)$ respectively, 
see Eq.~(\ref{eq:friction_functions}). We used numerical
methods \cite{clewley2012hybrid} to obtain the respective
full bifurcation diagrams, with solid/dashed lines denoting 
stable/unstable fixpoints and limit cycles. The corresponding 
flow in phase space is illustrated in Fig.~\ref{fig:bifucations_flow}.

For negative $\mu_1$ values the two fixpoints $(\pm1,0)$ are stable,
for the case of $f_1(V)$ and $f_3(V)$, and stable limit cycles evolve 
via two supercritical Hopf-bifurcation. For $f_2(V)$, on the other side,
a sub-critical Hopf bifurcation is observed at $\mu_1=0$.
The respective stable/unstable limit cycles merge for $f_1(V)$ and 
$f_2(V)$ in a homoclinic bifurcation, whereas a more complex 
bifurcation diagram emerges for $f_3(V)$. Saddle node bifurcation
of limit cycles are present for both $f_2(V)$ and $f_3(V)$.

%%%%%%%%%%%%%%%%%%%%%%%%%%%%%%%%%%%%%%%%%%%%%%%%%%%%%%%
\begin{figure}[t!]
\centering
\includegraphics[width=0.8\textwidth]{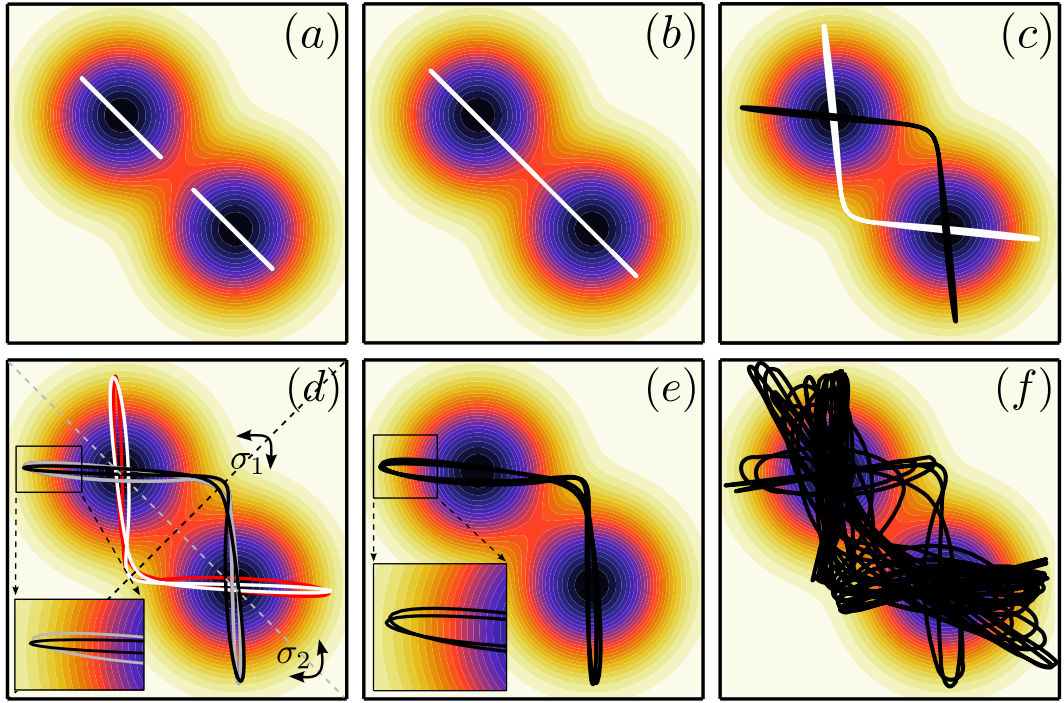}
\caption{Stable limit cycles and chaotic orbits of the Li\'enard 
prototype system (\ref{eq:prototype_system}) in a 
two-dimensional symmetric double well potential 
$V(\mathbf{x})$ (color coded, as defined by Eq.~(\ref{eq:V_x_1_2}))
and with a linear friction term 
$f_1(V(\mathbf{x})) = 0.5(\mu_1-V(\mathbf{x}))$. The bifurcation
parameter $\mu_1$ is $0.1, 0.15, 0.25, 0.265, 0.2698, 0.3$ from $(a)$ 
to $(f)$. In $(d)$ the four limit cycles can be mapped into 
each other by using the symmetry operations $\sigma_{1,2}$
or $\sigma_{3,4}$, as discussed in the Methods section.
For $(e)$ only a single of the four stable 
limit cycles is shown. This needs to circle the 
two potential minima twice in order to retrace itself. In $(f)$ 
an example of a chaotic trajectory is given.
}
\label{fig:double_2D_potential_well_orbits}
\end{figure}
%%%%%%%%%%%%%%%%%%%%%%%%%%%%%%%%%%%%%%%%%%%%%%%%%%%%%%%

%------------------------------------------------------
\subsection*{Chaos via period doubling of limit cycles}
\label{sec:chaos}
%------------------------------------------------------

\begin{figure}[t!]
\centering
\includegraphics[height=0.6\textwidth]{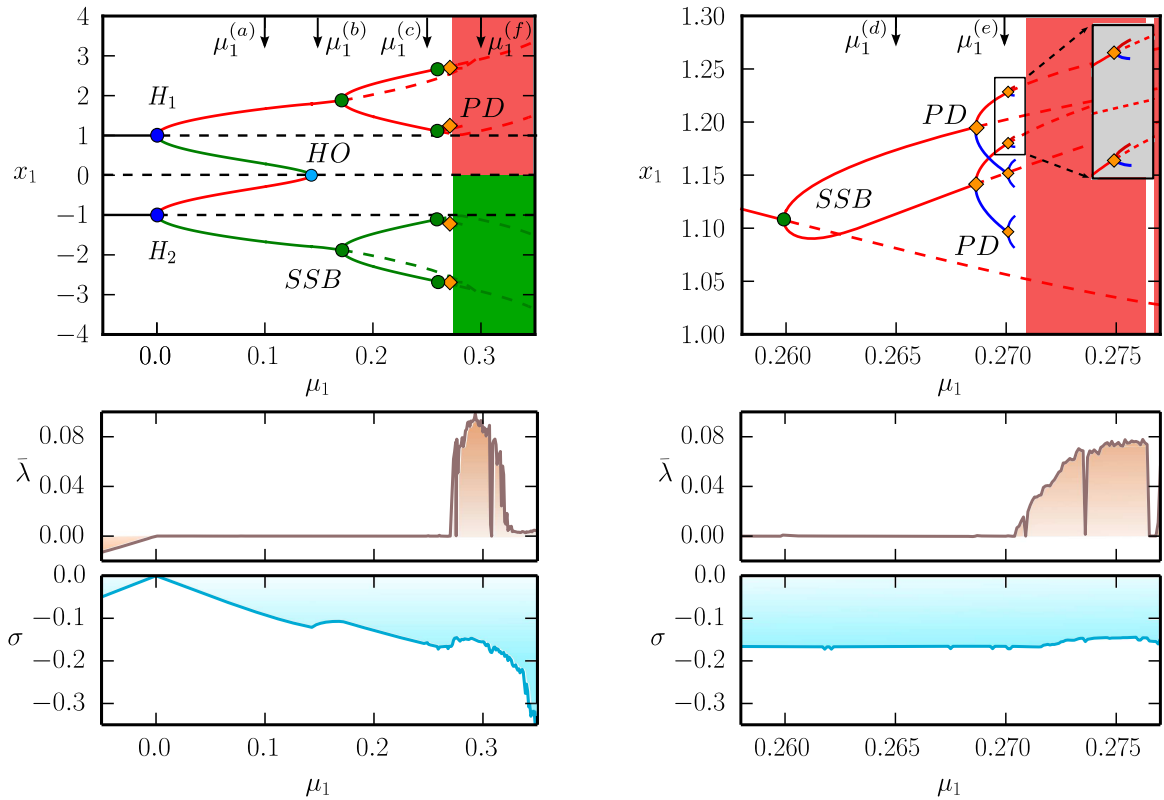}
\caption{\textit{Top row}: The numerically obtained bifurcation diagram 
for a two dimensional prototype Li\'enard system with symmetric 
double-well potential and a linear friction force, as for 
Fig.~\ref{fig:double_2D_potential_well_orbits}. The second branches 
of limit cycles emerging from the two destabilized minima 
(see Supplementary Fig.~S1) are not shown here. 
One observes Hopf and homoclinic bifurcations (H and HO), branching
of limit cycles via spontaneous symmetry breaking and period 
doubling (SSB and PD), as well a transition to chaos for 
$\mu>0.2705$. The red/green lines indicate the maximal/minimal $x_1$-values 
of the respective limit cycles. The blue curve is the second 
$x_1$-maxima after period doubling. The right diagram represents 
a zoom-in of the transition to a chaotic region, indicated by the 
shaded green and red areas. Only the first two period doubling 
bifurcations are shown.
\textit{Bottom row}: The average Lyapunov exponent $\bar{\lambda}$ 
and the contraction rate $\sigma$, calculated as described in 
the Methods section, for the corresponding $\mu_1$ parameter intervals. 
For the left figure $\Delta \mu_1=0.001$ parameter stepsize was used. 
Increasing the resolution more and more periodic windows 
(with $\bar{\lambda}=0$) become visible, as shown on the right 
plot, where $\Delta \mu_1$ is decreased with a factor of ten.}
\label{fig:double_2D_potential_well_bifurcations}
\end{figure}

We consider now a prototype system (\ref{eq:prototype_system})
with a two-dimensional symmetric potential field
$V(\mathbf{x})$, 
\begin{equation}
V(\mathbf{x}) = g(\mathbf{x}-\mathbf{x}_1) g(\mathbf{x}-\mathbf{x}_2),
\quad\quad g(\mathbf{z}) = \tanh(4\mathbf{z}^2/9)~,
\label{eq:V_x_1_2}
\end{equation}
as defined in (\ref{eq:V_predefined}), having two minima 
$\mathbf{x}_{1,2}=\pm(1,-1)$, and a linear friction term 
$f_1(V)=0.5(\mu-V)$. Both diagonals in the $(x_1,x_2)$ are 
symmetries of the system, as discussed in the Methods section.
In Fig.~\ref{fig:double_2D_potential_well_orbits} we present 
examples of stable limit cycles and of a chaotic trajectory,
as projected to the $(x_1,x_2)$ plane. 
In Fig.~\ref{fig:double_2D_potential_well_bifurcations}
the corresponding bifurcation diagram is presented, 
which shows Hopf bifurcations (H), homoclinic bifurcations (HO), 
branching of limit cycles via spontaneous symmetry breaking (SSB), 
period doubling of limit cycles (PD) and a transition to chaotic 
behavior:
\begin{description}
\item[H] At $\mu_1^{(H)}=0$ the two potential minima become unstable,
  just as for the one-dimensional spatial system presented in
  Fig.~\ref{fig:bifucations_V_diagrams}, resulting in two equivalent
  super-critical Hopf bifurcations. We note that, as a result of 
  the symmetric potential function (\ref{eq:V_x_1_2}), a second branch 
  of limit cycles is created by the two Hopf bifurcations
  (see the discussion in the Methods section and the Supplementary 
  Information). However, 
  since in the parameter region of interest these limit cycles are 
  mostly unstable, we have not investigated them in detail.
\item[HO] At $\mu_1^{(HO)}\approx0.143$ the limit cycles merge,
  as in 
  Fig.~\ref{fig:double_2D_potential_well_orbits}(a)~$\to$~(b), 
  in a homoclinic transition. The limit cycle stays, however, 
  exactly on the diagonal $x_1+x_2=0$.
\item[SSB] At the first branching point of limit cycles,
  $\mu_1^{(SSB)}\approx0.171$ the symmetry with respect to the 
  diagonal $(1,-1)$ is spontaneously broken, as in 
  Fig.~\ref{fig:double_2D_potential_well_orbits}(b)~$\to$~(c), 
  with the two limit cycles still being symmetric with respect to 
  the $(1,1)$ diagonal. The latter symmetry is broken at the second 
  branching point $\mu_1^{(SSB)}\approx0.260$, as in
  Fig.~\ref{fig:double_2D_potential_well_orbits}(c)~$\to$~(d), 
  with four symmetry related limit cycles being stable. 
\item[PD] For larger values of the bifurcation parameter $\mu_1$
  a series of period-doubling of limit cycles is observed, with
  the first occurring at $\mu_1^{(PD)}\approx0.268$, as in
  Fig.~\ref{fig:double_2D_potential_well_orbits}(d)~$\to$~(e). 
  The next period-doubling transition occurs at
  $\mu_1^{(PD)}\approx0.270$, as shown in
  Fig.~\ref{fig:double_2D_potential_well_bifurcations}.
\end{description}
For reference we note, that the saddle of the potential is 
located at $V(0,0)=0.505$, viz at a substantially larger 
value.

For $\mu_1>\mu_1^{(chaos)}\approx 0.2705$ we observe 
seemingly chaotic trajectories, as illustrated in
Fig.~\ref{fig:double_2D_potential_well_orbits}(f). 
Studying the transition to chaos is not the subject of
the present investigation and we leave it to future work.
We presume however, that the transition occurs via an 
accumulation of an infinite number of of period-doubling
transitions of limit cycles, similar to the ones observed 
for the Lorenz system \cite{robbins1979periodic} and for the
R\"ossler attractor \cite{rossler1976equation,gardini1985hopf}.

%%%%%%%%%%%%%%%%%%%%%%%%%%%%%%%%%%%%%%%%%%%%%%%%%%%%%%%
%%%%%%%%%%%%%%%%%%%%%%%%%%%%%%%%%%%%%%%%%%%%%%%%%%%%%%%

Our prototype system (\ref{eq:prototype_system}) is not
generically dissipative. We have evaluated the
average contraction rate $\sigma$, as defined
by (\ref{eq:sigma_def}) in the Methods section,
and present the results in 
Fig.~\ref{fig:double_2D_potential_well_bifurcations}.
Phase space contracts trivially along the attracting 
limit cycles, but also, on the average, in the chaotic 
region, where the average Lyapunov exponent $\bar{\lambda}$
becomes positive. $\bar{\lambda}$ is negative also
for $\mu_1<0$, when only stable fixpoints are present,
vanishing for intermediate values of $\mu_1$, when
stable limit cycles are present. The later is due to the 
fact, see Fig.~\ref{fig:dpw17_delta_r-time_example} and 
the corresponding Methods section, that
two initially close trajectories will generically flow to
the same limit cycle with the relative distance becoming
then constant.

\begin{figure}[t]
\centering
\includegraphics[height=0.3\textwidth]{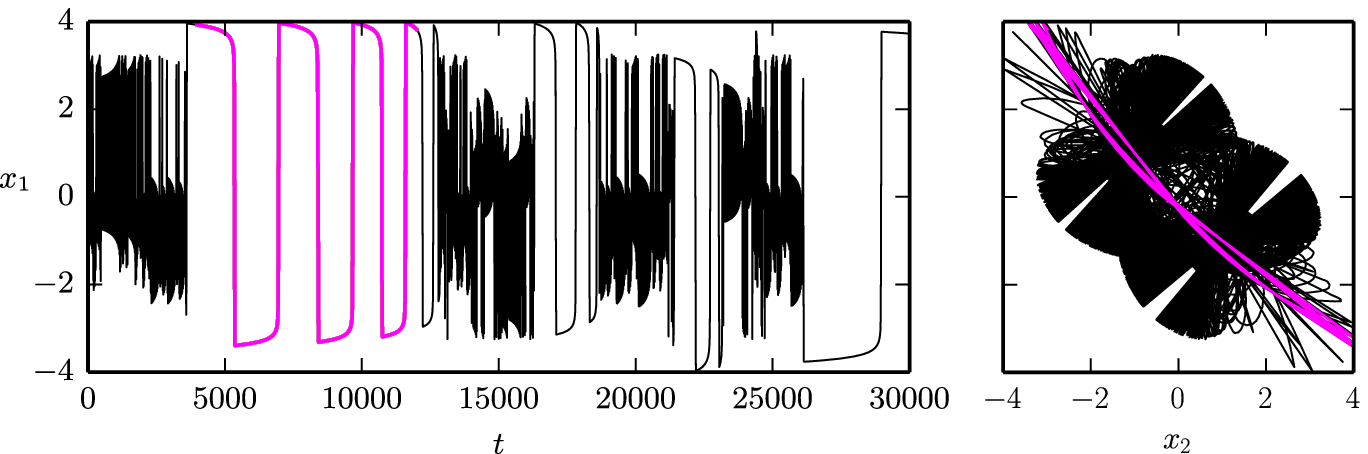}
\caption{\textit{Left}: Quasi-periodic windows with different 
time scales in the time-series plot of $x_1$  
variable for $\mu_1=0.34$.
\textit{Right}: Phase space plot of the trajectory 
projected to the $(x_1,x_2)$ plane for the same parameter. 
The magenta curve corresponds to the $t_p=[4\cdot 10^3,12\cdot10^3]$ 
quasi-periodic time interval, denoted with the same color 
in the $x_1(t)$ plot.
}
\label{fig:double_2D_potential_well_transient_chaos}
\end{figure}

For larger values of $\mu_1>0.322$ the chaotic
region transients into a phase of intermittent chaos
as illustrated in
Fig.~\ref{fig:double_2D_potential_well_transient_chaos},
in which an extended quasi-regular flow along the $(-1,1)$
diagonal is interseeded by a roughly perpendicular bursting
flow. This behavior is, to a certain extend, reminiscent to
a scenario of intermittent chaos \cite{lai1996symmetry},
in which a strange attractor is embedded in a higher-dimensional
space with partly unstable directions. We have, however, not
investigated the observed intermittent dynamics in detail.

%%%%%%%%%%%%%%%%%%%%%%%%%%%%%%%%%%%%%%%%%%%%%%%%%%%%%%%
%%%%%%%%%%%%%%%%%%%%%%%%%%%%%%%%%%%%%%%%%%%%%%%%%%%%%%%

%%%%%%%%%%%%%%%%%%%%%%%%%%%%%%%%%%%%%%%%%%%%%%%%
\section*{Discussion}
%%%%%%%%%%%%%%%%%%%%%%%%%%%%%%%%%%%%%%%%%%%%%%%%

We have proposed and discussed a prototype dynamical
system (\ref{eq:prototype_system}) in which the
friction forces $\propto f(V)$ depend functionally 
only on the mechanical potential $V(\mathbf{x})$. 
We have shown, that complex cascade of limit cycle
bifurcation can be obtained even for two dimensional
phase spaces when the friction function $f(V)$ alternates
between regions of energy uptake and dissipation.

We have also introduced a generic class of potential
functions (\ref{eq:V_predefined}) which allow to define,
in a relative straightforward manner, mechanical potentials 
with an arbitrary number of local minima with varying
depth. Any other potential could be however used. For
example one could study the biquadratic version
\begin{equation}
V(\mathbf{x}) \ \to\ 
\big(\mathbf{x}-\mathbf{x}_1\big)^2
\big(\mathbf{x}-\mathbf{x}_2\big)^2
\label{eq:V_x_1_2_bi_quadratic}
\end{equation}
of the potential (\ref{eq:V_x_1_2}) used in our 
study of chaotic behavior with the prototype 
system (\ref{eq:prototype_system}). We did not
study in detail the bifurcation diagram for the potential 
function (\ref{eq:V_x_1_2_bi_quadratic}), presume
however, that it would be similar to the results
presented in Fig.~\ref{fig:double_2D_potential_well_orbits}
and \ref{fig:double_2D_potential_well_bifurcations},
having the same underlying driving mechanism in terms
of a linear friction function $f(V)\propto(\mu_1-V)$.

We have shown that a simple double-well prototype
system with two spatial dimensions (and with a four-dimensional 
phase space) shows both symmetry induced bifurcations
of limit cycles together with a period-doubling of
limit cycle transition to chaotic behavior.

As a future perspective we note, that by changing the 
depth of the minima, one could control the order 
in which the fixpoints are going be destabilized, 
which might lead to other interesting phenomena. 
Adding an extra (maybe slow) dynamics to the positions 
or heights of the minima, the metadynamics of the attractors 
\cite{gros2014attractor}
may also be considered. In this case the $p_n$ parameters 
should be recalculated in each time-step using 
the self-consistent equations (\ref{eq:V_predefined_p_n}), 
which proved to be fast enough for practical purposes. 

Models, for which the equations of motion are derived 
from higher order principles, provide promising 
results for the understanding many different phenomena, 
such as the optimization hardness of boolean 
satisfiability problems \cite{ercsey-ravasz2011optimization} 
or the complex dynamics of biological neural networks 
\cite{markovic2010self,markovic2012intrinsic}.
Generally, these methods involve the construction of 
a generating functional, such as the cost function or energy
functional
\cite{intrator1992objective,triesch2005gradient,friston2010free,prokopenko2013guided},
with the dynamics of the system being defined by a gradient decent 
rule. When all equations are derived from the same generating 
functional, the system corresponds mathematically to 
a gradient system for which the asymptotic behavior 
is determined by the stable fixpoints (nodes). 
Thus they can not produce limit cycles or any 
oscillatory behavior. To by-pass this problem, 
usually additional equations of motions are defined, 
derived either from a second generating functional 
to produce objective function stress 
\cite{linkerhand2013generating,gros2014generating}
or from other considerations.
In our model, the system has an inherent inertia, 
which leads to damped oscillations around equilibria
(minima of the potential) in the presence of 
dissipation. By creating regions of
antidissipation, stable oscillatory dynamics and 
chaotic behavior can also be achieved.
Considering nonsymmetric, higher dimensional potential 
functions we expect to find an even richer set 
of dynamical behaviors (see the Methods section), 
a scenario worth be investigated in the future.

Finally, we note that the concepts 
of dynamical systems theory, such as attractors, 
slow points and bifurcations have been used recently 
to understand phenomena of surprisingly diverse 
fields. For examples of some relevant applications 
we mention here the modeling of birdsong 
\cite{sitt2008dynamical}, migraine 
\cite{dahlem2013migraine} dynamics and
the control mechanisms developed 
for the movements of humanoid robots 
\cite{ernesti2012encoding}.
Hence, we believe that prototype systems 
which allow the construction of models with 
predefined attractors in an intuitive manner 
could offer a useful tool for understanding
the behavior of interesting interdisciplinary problems.

%%%%%%%%%%%%%%%%%%%%%%%%%%%%%%%%%%%%%%%%%%%%%%%%
\section*{Methods}
\label{sec:methods}
%%%%%%%%%%%%%%%%%%%%%%%%%%%%%%%%%%%%%%%%%%%%%%%%

%------------------------------------------------------
\subsection*{Hopf bifurcations in the prototype system}
\label{sect:Hopf}
%------------------------------------------------------

\subsubsection*{2-dimensional prototype systems}

The local maxima of the potential function, i.e.\ where 
$V''(x^*)<0$, which are saddle points, separate the phase 
plane into different attraction domains with their stable 
manifold. Local minima with $V''(x^*)>0$ become, on the 
other hand, repelling focuses as a result of an 
Andronov-Hopf bifurcation, when dissipation changes to 
antidissipation in their neighborhood, having a simple 
pair of purely imaginary eigenvalues
$\lambda_{1,2}=\pm i\sqrt{V''(x^*)}$.

\subsubsection*{4-dimensional prototype systems}

Analogously, the 
\begin{equation*}
\mathbf{q}^*=(x_1^*,x_2^*,y_1^*,y_2^*),
\qquad\quad y^*_{1,2}=0,
\qquad\quad 
\left.\frac{\partial V(x_1,x_2)}{\partial x_{1,2}}\right\vert_{x_1^*,x_2^*} = 0
\end{equation*}
fixpoints of the 4-dimensional prototype systems 
(\ref{eq:prototype_system}) correspond to critical points 
of the $V(x_1,x_2)$ potential function. Classification of the 
local minima and saddle critical points with respect to their 
stability can be achieved by evaluating the eigenvalues of the 
Jacobian of the system in terms of the Hessian of the potential
function:
\begin{equation}
J(\mathbf{q}^*) = 
\begin{pmatrix}
0 & 0 & 1 & 0 \\
0 & 0 & 0 & 1 \\
d_1 &  c & a & 0 \\
c & d_2 & 0 & a
\end{pmatrix},\qquad\quad
H(x^*_1,x^*_2)=
\begin{pmatrix}
-d_1 & -c \\
-c & -d_2
\end{pmatrix}
\end{equation}
where we have defined with
\begin{equation}
a = f(V(x^*_1,x^*_2)), \qquad\quad
c = - \left.\frac{\partial^2 V}{\partial x_1 \partial x_2}\right\vert_{x_1^*,x_2^*}, \qquad\quad
d_{1,2} = - \left.\frac{\partial^2 V}{\partial x_{1,2}^2}\right\vert_{x_1^*,x_2^*}
\end{equation}
the friction term and the second order partial derivatives of the 
potential at the respective critical points. Defining with
\begin{equation}
\gamma_{\pm} = \frac{1}{2}\left(-(d_1+d_2)\pm \sqrt{(d_1+d_2)^2 - 4(d_1 d_2 -c^2)}\right),
\qquad\quad \gamma_{\pm} \in \mathbb{R}
\label{eq:gamma_pm}
\end{equation}
we can express the eigenvalues of the Jacobian $J(\mathbf{q}^*)$ as
\begin{equation}
\lambda_{1,2,3,4}=\frac{1}{2}\left(a \pm \sqrt{a^2-4\gamma_{\pm}}\right).
\end{equation}
For general potential functions the local minima, 
defined by the $\Delta=\det(H)=d_1d_2-c^2>0$ 
and $\rho=\tr(H)=-(d_1+d2)<0$ conditions  
(or equivalently by $\gamma_{\pm}>0$), 
undergo a Hopf bifurcation, 
when the $f(V)$ friction term changes sign, i.e.:
\begin{equation}
a = f(V(x_1^*,x_2^*)) = 0 \qquad\Rightarrow\qquad \lambda_{1,2,3,4}=\pm i\sqrt{\gamma_{\pm}}.
\end{equation}
However, saddles of the potential function, 
i.e. $\Delta=\det(H)<0$, are saddle type
fixpoints of the dynamical system, 
having always a $\lambda_1$ positive eigenvalue,
as $\gamma_{+}>0$ and $\gamma_{-}<0$.

Here we note that in case of the potential function 
(\ref{eq:V_x_1_2}), due to the symmetries one gets  
a double pair of imaginary eigenvalues, since
$d_1=d_2$ and $c=0$, and thus Eq.~(\ref{eq:gamma_pm}) 
yields $\gamma_{+}=\gamma_{-}$.
This results in a second branch of limit cycle 
solutions, not investigated in this paper, 
emerging from the Hopf-point.

\subsubsection*{2d-dimensional prototype systems}

For arbitrary dimensions $d$ one can express the
Jacobian in terms of block matrices:
\begin{equation}
J(\mathbf{q}^*) = 
\begin{pmatrix}
O_d & I_d \\
- H_d & a I_d
\end{pmatrix},
\end{equation}
where $a=f(V)$ and where $O_d$ and $I_d$ are 
the $d$-dimensional zero and identity matrices.
$H_d=(H_{i,j}(\mathbf{x}^*))=\left(\left.\frac{\partial^2 V}
{\partial x_i\partial x_j}\right\vert_{\mathbf{x}^*}\right)$ 
is the Hessian matrix of the $V(\mathbf{q})$ potential, 
evaluated for the respective 
$\mathbf{q}^*=(\mathbf{x}^*,\mathbf{y}^*)$ fixpoint.

To determine the eigenvalues of the Jacobian 
one has to solve the equation:
\begin{equation}
\det(J-\lambda I_d) = 
\begin{vmatrix}
- \lambda I_d & I_d \\
- H_d & (a-\lambda)I_d
\end{vmatrix}=
\det(-\lambda(a-\lambda)I_d + H_d)=0,
\end{equation}
where we used the properties of square 
block matrices. By introducing
$\gamma=\lambda(a-\lambda)$ on finds
with
\begin{equation}
\det(J-\lambda I_d) = \det(H_d -\gamma I_d) 
= \prod_{i=1}^d(\gamma-\gamma_i)=0
\end{equation}
that the $2d$ eigenvalues $\lambda$ of the Jacobian can be
expressed in terms of the $d$ eigenvalues 
$\gamma_i$ of the Hessian matrix and hence
\begin{equation}
\lambda_i^{\pm} = \frac{1}{2}\left(a\pm\sqrt{a^2-4\gamma_i}\right).
\end{equation}
Consequently, at the local minima of the potential, 
i.e.\ when $\gamma_i>0$, a Hopf-bifurcation occurs, 
with $\lambda_i^{\pm}=\pm i\sqrt{\gamma_i}$, 
when the friction term $a=f(V)$ changes sign.
For general potential functions this might lead to    
the birth of higher dimensional tori 
or several branches of limit cycle bifurcations.

%------------------------------------------------------
\subsection*{Symmetries of the 4-dimensional system}
\label{sect:4D_2D}
%------------------------------------------------------
The results shown in 
Figs.~\ref{fig:double_2D_potential_well_orbits},
\ref{fig:double_2D_potential_well_bifurcations} and
\ref{fig:double_2D_potential_well_transient_chaos}
are for a 4-dimensional prototype systems 
(\ref{eq:prototype_system}) with a linear friction
force $f_1(V)$, as defined in (\ref{eq:friction_functions}),
and a mechanical potential $V(\mathbf{x})$ given by 
(\ref{eq:V_x_1_2}). The minima $V(\mathbf{x}_{1,2})=0$
of the potential, viz.\
$\mathbf{x_1} = (+1,-1)$ and
$\mathbf{x_2} = (-1,+1)$ are connected by the symmetry
operations
\begin{equation}
\sigma_{1,2}= 
\begin{pmatrix}
0 & \pm1 & 0 & 0\\
\pm1 & 0 & 0 & 0\\
0 & 0 & 1 & 0\\
0 & 0 & 0 & 1
\end{pmatrix},\qquad\quad
\sigma_{3,4}= 
\begin{pmatrix}
0 & \pm1 & 0 & 0\\
\pm1 & 0 & 0 & 0\\
0 & 0 & 0 & \pm1\\
0 & 0 & \pm1 & 0
\end{pmatrix}
\label{eq:sigma}
\end{equation}
of the system. Thus, if $(x_1,x_2,y_1,y_2)$ is a 
solutions, then 
\begin{equation}
\begin{pmatrix}
x_1^\prime \\ x_2^\prime \\ y_1^\prime \\ y_2^\prime
\end{pmatrix} = \sigma_{1,2}
\begin{pmatrix}
x_1 \\ x_2 \\ y_1 \\ y_2
\end{pmatrix},\qquad\quad
\begin{pmatrix}
x_1^{\prime\prime}\\ x_2^{\prime\prime} \\
y_1^{\prime\prime} \\ y_2^{\prime\prime}
\end{pmatrix} = \sigma_{3,4}
\begin{pmatrix}
x_1 \\ x_2 \\ y_1 \\ y_2
\end{pmatrix}
\end{equation}
are also solutions. 

%%%%%%%%%%%%%%%%%%%%%%%%%%%%%%%%%%%%%%%%%%%%%%%%%%%%%%%
\begin{figure}[t!]
\centering
\includegraphics[width=0.43\textwidth]{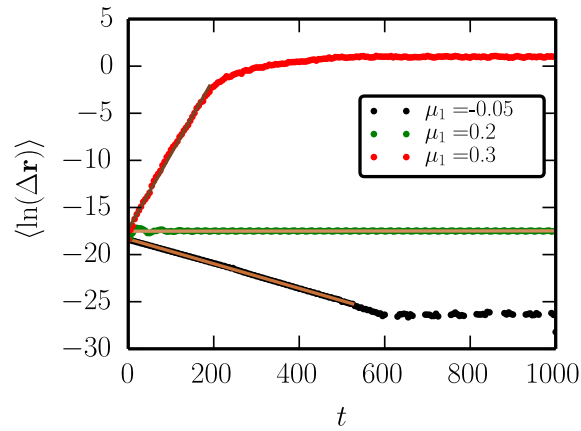}
\caption{The logarithmic growth rate 
$\langle\ln(\Delta\mathbf{r})\rangle$ averaged for $100$ 
random initial conditions as a function of time for 
three qualitatively different types of dynamics: spiraling 
into a fixpoint ($\mu_1=-0.05$), limit cycle oscillations 
($\mu_1=0.2$) and chaotic behavior ($\mu_1=0.3$).
Brown lines correspond to the best linear regression.
In the first and last case, the line is fitted 
only to the first part of the trajectory.
The dashed line indicates that the distance of the point 
pairs has reached the maximal accuracy of the integrator.
}
\label{fig:dpw17_delta_r-time_example}
\end{figure}
%%%%%%%%%%%%%%%%%%%%%%%%%%%%%%%%%%%%%%%%%%%%%%%%%%%%%%%

%----------------------------------------------
\subsection*{Lyapunov exponent and contraction rate}
\label{subsect:methods-Lyapunov_exponent_contraction_rate}
%----------------------------------------------

The local Lyapunov exponent $\lambda$ is determined from
the growth rate of the distance 
$\Delta\mathbf{r}(t)=\Delta\mathbf{r}_0e^{\lambda t}$, 
between point pairs with an initial displacement, which
we have taken to be $\Delta\mathbf{r}_0=10^{-8}$. 
Measuring the Lyapunov exponent was started after a
transient of $t_{tr}=1.5\cdot10^4$. Considering
100 random initial conditions the 
average Lyapunov exponent $\bar\lambda$ is 
then given by the slope of the initial linear 
part of the $\langle\ln(\Delta\mathbf{r})\rangle$ curve 
(as given by the brown lines in 
Fig.~\ref{fig:dpw17_delta_r-time_example}).

The contraction rate $\sigma$, is defined as the average 
of local contraction rates along a set of 
trajectories $\Gamma$ for different initial conditions:
\begin{equation}
\sigma\ =\ \left\langle\frac{1}{L}
\int_{\Gamma} \nabla\cdot\mathbf{f}\, ds
\right\rangle,
\label{eq:sigma_def}
\end{equation}
where $L=\int_{\Gamma}ds$ is the length of the trajectory and 
$\mathbf{f}$ is the flow, viz the right-hand side of the 
evolution equations (\ref{eq:prototype_system}).
$\sigma$ is negative for dissipative systems,
in which the phase space contracts
\cite{gros2013complex,chow1994normal}.

%%%%%%%%%%%%%%%%%%%%%%%%%%%%%%%%%%%%%%%%%%%%%%%%%%%%%%
%\bibliography{generalised_lienard_system}

%%%%%%%%%%%%%%%%%%%%%%%%%%%%%%%%%%%%%%%%%%%%%%%%%%%%%%%

%%%%%%%%%%%%%%%%%%%%%%%%%%%%%%%%%%%%%%%%%%%%%%%%%%%%%%%
\section*{Acknowledgements}
%%%%%%%%%%%%%%%%%%%%%%%%%%%%%%%%%%%%%%%%%%%%%%%%%%%%%%%
The work of B.S. was supported by the European Union 
and the State of Hungary, co-financed by the European
Social Fund in the framework of 
T\'AMOP 4.2.4.A/2-11-1-2012-0001 National Excellence Program.

\end{document}